\documentclass[twocolumn,
prl,aps,showpacs,nofootinbib]{revtex4}
\usepackage{epsfig}
\usepackage{amsmath,amssymb}
\renewcommand{\vec}[1]{\boldsymbol{{\bf #1}}}
\begin{document}

\title{
Generation of harmonics by a focused laser beam in
the vacuum}
\author{A.M. Fedotov\thanks{E-mail: fedotov@cea.ru},
N.B. Narozhny\thanks{E-mail: narozhny@theor.mephi.ru}}
\address{Moscow Engineering Physics Institute,
115409 Moscow, Russia}

\begin{abstract}
We consider generation of odd harmonics by a super strong focused
laser beam in
the vacuum. The process occurs due to the plural
light-by-light scattering effect. In the leading order of
perturbation theory, generation of $(2k+1)$th harmonic is described
by a loop diagram with $(2k+2)$ external incoming, and two outgoing
legs. A frequency of the beam is assumed to be much smaller than the
Compton frequency, so that the approximation of a constant uniform
electromagnetic field is valid locally. Analytical expressions for
angular distribution of generated photons, as well as for their
total emission rate are obtained in the leading order of
perturbation theory. Influence of higher-order diagrams is studied
numerically using the formalism of Intense Field QED. It is shown
that the process may become observable for the beam intensity of the
order of $10^{27}\rm{W/cm}^2$.
\end{abstract}
 \maketitle

In a recent paper \cite{Mourou} T. Tajima and G. Mourou suggested a
path to reach an extremely high intensity level $10^{26-28}$W/cm$^2$
already in the coming decade, taking advantage of the megajoule
facilities, see also Refs.~\cite{Shen,Bulanov}. The field strength
for such lasers will be very close to the characteristic QED value
$E_S=m^2c^3/e\hbar=1.32\cdot 10^{16}\rm{V/cm}$, and thus nonlinear
QED vacuum polarization effects will become measurable. Some of them
have been already studied in literature. Among those, harmonic
generation by an intense laser beam propagating in an external
magnetic field \cite{DingAndCaplan}, and by two colliding beams in
the vacuum \cite{HG_st_w}. Recently the experimental feasibility of
the light-by-light stimulated scattering via three laser beams has
been studied in Ref.~\cite{3beams}. Pair creation by a focused laser
pulse in the vacuum was considered in Ref.~\cite{Nar1}, and by two
colliding pulses in Ref.~\cite{Nar2}. Other references could be
found in the recent review \cite{rev}. In this letter, we consider
the effect of odd harmonics generation by a strong focused laser
beam in the vacuum. Clearly, the probability of this process is
smaller compared to an analogous light-by-light scattering effect in
a combination of laser beams \cite{3beams}. However it is of great
importance, since the superhigh laser intensities can be achieved
only in focused and very short laser pulses, and hence the harmonics
generation effect will necessarily be present at any facility
producing pulses with the peak field strength of the order of $E_S$.
\begin{figure}[hb]
\epsfxsize8cm\epsffile{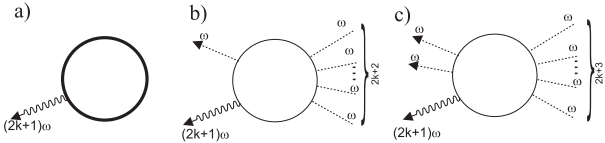} \caption{\small a) The exact
diagram for generation of $l=(2k+1)$-th harmonic (a thick line
corresponds to the exact electron propagator); b), c) The diagrams
for generation of $l=(2k+1)$-th harmonic in the two leading orders
of perturbation theory. Dashed lines represent laser photons of
frequency $\omega$.} \label{diagrams}
\end{figure}

The process under consideration is described by the Feynman diagram
depicted in Fig.~\ref{diagrams}a. In the two leading orders of
perturbation theory the effect is represented by diagrams in
Fig.~\ref{diagrams}b,c. The diagrams without outgoing laser photons
do not contribute since they would correspond to creation of an
extraneous photon by a group of laser photons propagating in one
direction. As is well known, such photons do not interact.
For the fields weakly varying at distances of the order of the
Compton length $l_C$, and time intervals $l_C/c$, nonlinear vacuum
polarization effects can be described using the Heisenberg-Euler
radiative correction to the electromagnetic field Lagrangian
density \cite{HE,Weisskopf,Schwinger}, which can be represented by
the following asymptotic expansion
\begin{equation}\label{Lagr_ren}
{\cal L}'= \alpha\left\{\frac{4{\cal F}^2+7{\cal
G}^2}{360\pi^2E_S^2}
+\frac{{\cal F}\left(8{\cal F}^2+13{\cal
G}^2\right)}{630\pi^2E_S^4}+ \ldots\right\},
\end{equation}
$\alpha=e^2/\hbar c$ is the fine structure constant, ${\cal
F}=(\vec{E}^2-\vec{H}^2)/2$ and ${\cal G}=\vec{E}\cdot\vec{H}$ are
the invariants of the electromagnetic field\footnote{From now on,
the natural units $\hbar=c=1$ are used.}. The relation $\lambda\gg
l_C$, where $\lambda$ is the wave length of laser radiation, is
valid for any of existing or projected lasers
(with a possible exception of
$\gamma$, or hard X-ray lasers). Thus the Lagrangian density
(\ref{Lagr_ren}) can be evidently applied for description of focused
laser fields.

The Lagrangian density (\ref{Lagr_ren}) generates effective vacuum
polarization $\vec{P}=\partial {\cal L'}/\partial\vec{E}$ and
magnetization $\vec{M}=-\partial {\cal L'}/\partial\vec{H}$ and the
effective vacuum charge and current densities
\begin{equation}\label{vac_curr}
\rho=-\nabla\cdot\vec{P},\quad \vec{j}=\left(\frac{\partial
\vec{P}}{\partial t}-\nabla\times\vec{M}\right).
\end{equation}
Then the modified vacuum Maxwell equations in the Lorentz gauge are
reduced to the form
\begin{equation}\label{Maxwell}
\square A_{\mu}=4\pi j_{\mu}(\vec{E},\vec{H}).
\end{equation}

The vacuum current (\ref{vac_curr}) contains the fine structure
constant $\alpha$. Therefore we will look for a solution to
Eq.~(\ref{Maxwell}) (as well as the fields $\vec{E},\vec{H}$) in the
form of perturbative expansion, $A_{\mu}=A_{\mu}^{(0)}+A_{\mu}'$
($\vec{E}=\vec{E}^{(0)}+\vec{E}'$,
$\vec{H}=\vec{H}^{(0)}+\vec{H}'$), where the fields with zero index
obey the classical vacuum Maxwell equations. For the field in the
zeroth-order approximation we will use the model of a focused laser
beam developed in Ref.~\cite{NF1}. This model is based on an exact
solution of Maxwell equations and was successfully employed to
explain experimental results on scattering of relativistic electrons
by a focused laser beam \cite{NF2}, and to estimate the number of
pairs created by a focused laser beam in the vacuum
\cite{Nar1,Nar2}. A similar model was used in Ref.~\cite{Ind1}.

Following Ref.~\cite{NF1} we describe a focused laser
beam as a superposition of monochromatic plane waves of the same
frequency $\omega$ and with wave vectors lying inside a cone with
aperture angle $2\Delta$. The vector potential of the beam
propagating along the $z$ axis can be written then as
\begin{equation}\label{Pulse}
\vec{A}^{(0)}(\vec{r}_\perp,z,t)=\int\limits_{k_\perp<\omega}
d^2k_\perp \vec{a}(\vec{k}_\perp)
e^{i(\vec{k}_\perp\vec{r}_\perp+k_zz-\omega t)},
\end{equation}where
$$\vec{k}=\vec{k}_{\perp}+k_z\hat{\vec{z}},\quad
k_z=\sqrt{\omega^2-k_\perp^2},\quad\vec{k}\cdot\vec{a}(\vec{k}_\perp)=0.$$
The configuration of the beam is determined by the function
$\vec{a}(\vec{k}_\perp)$ and may vary strongly. However,
independently of the form of $\vec{a}(\vec{k}_\perp)$, any polarized
beam may be expressed as a superposition of $e$- and $h$-polarized
waves, i.e. the waves respectively with the vectors
$\vec{E}^{(0)}=i\omega\vec{A}^{(0)}$ and
$\vec{H}^{(0)}=\nabla\times\vec{A}^{(0)}$ perpendicular to the
propagation direction of the beam. We will choose the following form
for functions $\vec{a}^{e,h}(\vec{k}_\perp)$ for $e$- and $h$-
circularly polarized beams
\begin{equation}\label{Pulse1}
\vec{a}^{e(h)}(\vec{k}_\perp)=\frac{E_0^{(0)}k_\perp^2\big
(\vec{l}_{\perp}(\vec{k}) \cdot\vec{e}\big)
\vec{l}_{\perp(\parallel)}(\vec{k})}{4\sqrt{2}\pi\omega^5\Delta^4}\,
e^{-k_\perp^2/4\omega^2\Delta^2},
\end{equation}
where
$\vec{l}_{\perp}(\vec{k})=(\vec{k}\times\hat{\vec{z}})/k_{\perp}$,
$\vec{l}_{\parallel}(\vec{k})=(\vec{k}\times\vec{l}_{\perp})/\omega$
and $\vec{e}=(\hat{\vec{x}}\pm i\hat{\vec{y}})/\sqrt{2}$. The
resulting expressions for the fields can be found in
Refs.~\cite{NF1,Nar1}. They describe a focused beam with the focal
spot of the characteristic radius $R=1/\omega\Delta$  and the
diffractive length $L=1/\omega\Delta^2$. It will be important for us
later on in this Letter that the fields acquire the following
structure
\begin{equation}\label{f-str}
\vec{E}^{(0)},\vec{H}^{(0)}=E_0^{(0)}e^{-i\varphi}
\vec{F}(\vec{\xi},\chi),
\end{equation}
where $\varphi=\omega(t-z),\,\vec{\xi}=\vec{r_\perp}/R,\,\chi=z/L$.
The quantity $E_0^{(0)}$ is the amplitude of electric (magnetic)
field at the focal axis of the beam. It is worth noting that the
type of polarization of the beam is determined by the field
configuration near the focal axis (see Refs.~\cite{NF1,NF2} for
details).

For $\Delta\ll 1$ the integral in (\ref{Pulse}) can be evaluated
analytically. The formulas for the field invariants ${\cal F}$,
${\cal G}$ can be found in Ref.~\cite{Nar1}. Here we give explicit
expressions only for the invariants ${\cal F}$, ${\cal G}$ in the
focal plane $z=0$ ($\phi$ is the polar angle in that plane),
\begin{equation}\label{inv}
\begin{array}{l}\displaystyle
{\cal F}^e=E_0^2\Delta^2 e^{-2\xi^2}\left\{-8\xi^6+32\xi^4\right.\\
\\\displaystyle\left. +6\xi^2\cos[2(\omega t\mp\phi)]-28\xi^2+4\right\},\\
\\\displaystyle {\cal
G}^e=2E_0^2\Delta^2\xi^2e^{-2\xi^2}\sin[2(\omega t\mp\phi)],
\end{array}
\end{equation}
the upper (lower) sign corresponds to the clockwise-
(counterclockwise-) polarized waves. The fields in $e$- and
$h$-polarized waves are related as $\quad
\vec{E}^{(0)h}=\vec{H}^{(0)e}$, $ \vec{H}^{(0)h}=-\vec{E}^{(0)e}\,$
\cite{NF1}.
Thus, invariants for the $h$-wave are given by ${\cal F}^h=-{\cal
F}^e$, ${\cal G}^h=-{\cal G}^e$. It is clear that in both cases they
decrease exponentially away from the focal spot. Similarly, the
invariants decrease fast (in general as $\chi^{-6}$) away from the
focal plane. Therefore, the induced vacuum polarization charges and
currents (\ref{vac_curr}) are almost localized inside the focal
region with the volume $V_f\sim R^2L\sim 1/\omega^3\Delta^4$. For
small $\Delta$, the characteristic dimensions of this region are
essentially greater than the wavelength of the laser field. Hence,
space outside of the focal region constitutes a wave zone for the
radiation emitted by the vacuum polarization currents.

It is convenient to expand
the fields $\vec{E}'$, $\vec{H}'$ generated by vacuum currents in a
Fourier series, e.g., $\vec{H}'=\sum_l \vec{H}_l'(\vec{r}) e^{-i
\omega l t}$. It is well known, see, e.g., Ref.~\cite{LL}, that
Fourier components of magnetic strength of the radiation field in
the wave zone are given by
\begin{equation}\label{sol_Maxwell}
\vec{H}_l'(\vec{r})=\frac{i l\omega e^{i\omega
lr}}{rT}\int\limits_0^Tdt'\int\limits d^3r'e^{i l\omega(
t'-\vec{n}\vec{r}')}\,\vec{n}\times \vec{j}(\vec{r}',t'),
\end{equation}
$\vec{n}=\vec{r}/r$. Using Eqs.~(\ref{vac_curr}) we reduce
Eq.~(\ref{sol_Maxwell}) after integration by parts to the form
\begin{equation}\label{sol_Maxwell1}
\begin{array}{l}\displaystyle \vec{H}_l'=\frac{\omega^2l^2}{r} e^{i\omega
lr}\big(h_l^{(\perp)}\vec{l}_{\perp}(\vec{n})+h_l^{(\parallel)}
\vec{l}_{\parallel}(\vec{n})\big)\,,\\
\displaystyle h_l^{(i)}=\frac1{T}\int\limits_0^Tdt'\int\limits
d^3r'\,e^{i l\omega(
t'-\vec{n}\vec{r}')}\widetilde{h}_l^{(i)}(t',\vec{r}')\,,
\end{array}
\end{equation}
\begin{equation}\label{sol_Maxwell2}
\begin{array}{l}
\displaystyle
\widetilde{h}_l^{(i)}(t',\vec{r}')=\left(\frac{\partial {\cal
L'}}{\partial{\cal
G}}\right)_0\,\vec{l}_i(\vec{n})\cdot\left(\vec{E}^{(0)}+\vec{n}\times
\vec{H}^{(0)}\right) \\ \displaystyle -\left(\frac{\partial {\cal
L'}}{\partial{\cal
F}}\right)_0\,\vec{l}_i(\vec{n})\cdot\left(\vec{H}^{(0)}-\vec{n}\times
\vec{E}^{(0)}\right).
\end{array}
\end{equation}

The angular distribution of the radiation power at frequency $\omega
l$ is given by $dI_l=|\vec{H}_l'|^2r^2\,d\Omega_{\vec{k}}/2\pi$, see
Ref.~\cite{LL}. To obtain the rate $P_l$ of photon emission we
should divide this quantity by the photon frequency $\omega l$ and
then integrate it over the spatial angle,
\begin{equation}\label{result}
P_l=\frac{\omega^3l^3} {2\pi}\int\limits
d\Omega_{\vec{n}}\,(|h_l^{(\perp)}|^2+|h_l^{(\parallel)}|^2)\,.\end{equation}
We arrive at the same result computing the Feynman diagram
Fig.~\ref{diagrams}a where the bold line corresponds to the exact
electron propagator in a locally constant electromagnetic field
\cite{Schwinger}.

Expanding the vacuum current in a Fourier series, we represent the
probability of photon emission as a sum of partial probabilities,
each with its own conservation law corresponding to conversion of
some number $n$ of laser photons into $n'$ laser photons ($n'<n$)
and an extraneous one with the frequency $\omega'=l\omega$. It
immediately follows from the Furry theorem that the frequency
multiplier $l$ can be an odd number only, $l=2k+1$. This can be seen
also from Eq.~(\ref{result}), since the derivatives of the effective
Lagrangian (\ref{Lagr_ren}) with respect to the invariants are even
functions of electromagnetic field strengths.

Let us estimate now the rate of photon generation (\ref{result})
under assumption $\Delta\ll 1$. First we note, that due to
Eq.~(\ref{f-str}),  $\widetilde{h}_l^{(i)}(t',\vec{r}')$ is a
function of variables $\varphi', \xi', \phi', \chi'$. Passing on to
these variables, we write down Eq.~(\ref{sol_Maxwell2}) in the form
\begin{equation}\label{sol_Maxwell3}
\begin{array}{l}\displaystyle
h_l^{(i)}=\frac{V_f}{2\pi}\int\limits_0^{2\pi}d\varphi'\int\limits
\xi'd\xi'd\phi'd\chi'\,\widetilde{h}_l^{(i)}(\varphi',\vec{\xi}',\chi')\\
\displaystyle \times \exp\left\{i
l\varphi'-il\frac{\sin\theta}{\Delta}
\xi'\sin\phi'+il\frac{1-\cos\theta}{\Delta^2}\chi'\right\}\,,
\end{array}
\end{equation}where $\theta$ is the azimuthal
angle of the emitted photon,
$\cos\theta=(\vec{n}\cdot\hat{\vec{z}})$. As it can easily be seen
from Eq.~(\ref{Pulse1}), the effective values of the variable of
integration $k_{\perp}$ in the integral (\ref{Pulse}) are of the
order $k_\perp\sim\omega\Delta$ when $\Delta\ll 1$. Then, it can be
shown using Eqs.~(\ref{Pulse1}) and the definitions of the fields
$\vec{E}^{(0)},\,\vec{H}^{(0)}$ (\ref{Pulse}), that the combinations
$$\vec{l}_i(\vec{n})\cdot\left(\vec{H}^{(0)}-\vec{n}\times
\vec{E}^{(0)}\right), \quad
\vec{l}_i(\vec{n})\cdot\left(\vec{E}^{(0)}+\vec{n}\times
\vec{H}^{(0)}\right)\,, $$ which appear in Eq.~(\ref{sol_Maxwell2}),
are of the order $E_0^{(0)}\Delta^2$ for both $e$- and $h$-polarized
waves. Both invariants $\mathcal{F}$ and $\mathcal{G}$ are
$\sim(\Delta E_0^{(0)})^2$ in the focal region, see Eq.~(\ref{inv}).
Then, it follows from Eq.~(\ref{Lagr_ren}) that the derivatives
$\partial{\cal L}/\partial {\cal F}$, $\partial{\cal L}/\partial
{\cal G}$ are of the form $\alpha\,
g(\vartheta;\varphi',\vec{\xi}',\chi')$ near the focus, where $g$ is
a dimensionless function, and the dimensionless parameter
$\vartheta$ is defined as $\vartheta=\Delta E_0^{(0)}/E_S$. Thus the
coefficients $h_l^{(i)}$ in Eq.~(\ref{sol_Maxwell1}) acquire at
$\Delta\ll 1$ the form $h_l^{(i)}=\alpha
V_fE_0^{(0)}\Delta^2\mathrm{g}_{\,li}(v,\vartheta)$, where
$v=\theta/\Delta$. We have taken into account the fact that photons
are emitted basically forward with $\theta\lesssim\Delta$ when
$\Delta\ll 1$, as it can be seen from Eq.~(\ref{sol_Maxwell3}).
Substituting this estimate into Eq.~(\ref{result}) we obtain for the
number of photons generated per one period, $N_l=P_l T$:
\begin{equation}\label{main_res}
\begin{array}{l}\displaystyle
N_l=\omega^2l^3 \alpha^2(V_f)^2(E_0^{(0)}\Delta^2)^2\Delta^2\\
\displaystyle \times\sum\limits_i \int\limits_0^\infty dv\,v
|\mathrm{g}_{\,li}(v,\vartheta)|^2=\alpha
\left(\frac{m}{\omega\Delta}\right)^4\vartheta^2f_l(\vartheta).
\end{array}
\end{equation}
The quantity $(m/\omega\Delta)^4$ has the order of the ratio
$V_fT/l_C^4$ and is extremely large for optical lasers. For example
for a laser field with the wavelength $\lambda\sim 1{\rm \mu m}$ and
$\Delta=0.1$ it is  $\sim 10^{26}$. Therefore the number of
generated photons will become perceptible already at $\vartheta\ll
1$, i.e. at the laser field strength $E_0^{(0)}$  essentially less
than
$E_S$.

In the framework of the perturbation theory the leading contribution
to generation of the $l$-th harmonic ($l=2k+1$) arises from the
diagram with $l+1$ incoming and a single outgoing laser photon
lines, see Fig.~\ref{diagrams}b. Hence,
$f_l(\vartheta)=C_l\vartheta^{2l+2}$. The coefficients $C_l$ can be
computed analytically. For the first few harmonics one obtains
$C_3=6.380\cdot 10^{-4}$, $C_5=1.048\cdot 10^{-3}$, and
$C_7=3.060\cdot 10^{-3}$. Thus for the number of generated photons
at $\vartheta=10^{-2}$ we have $N_3\approx10$. For the laser field
with the wavelength $\lambda\sim 1{\rm \mu m}$ and $\Delta=0.1$ the
value $\vartheta=10^{-2}$ corresponds to the peak field strength in
the focus $E_0^{(0)}\approx0.1E_S$, or to laser intensity
$I\sim10^{-2}I_S$, where $I_S=(c/4\pi)E_S^2=4.65\cdot
10^{29}\rm{W/cm}^2$.

It is clear, that at $E_0^{(0)}\sim E_S$ and higher, the number of
generated photons becomes enormous. Therefore it makes sense to
compare the energy $W_T$ flowing through the focal spot during the
wave period $T$ with the energy of photons $W_T^{(\gamma)}$
generated during the same interval of time. The quantity $W_T$ can
be estimated as $W_T\sim (E_0^{(0)2}/4\pi)\pi R^2T\,,$ while
$W_T^{(\gamma)}=\sum\limits_l\omega lN_l$, with $N_l$ from
Eq.~(\ref{main_res}). Finally, we have $W_T^{(\gamma)}/W_T\sim
\alpha^2\sum\limits_llf_l(\vartheta)$. We see that this ratio
remains small at least for $\vartheta$ close to unity. This means
that exhaustion of the laser pulse due to harmonics generation
effect does not occur, as it happens in a pair creating laser pulse,
see Ref.~\cite{Nar1}.

\begin{figure}[h]
\epsfxsize6cm\epsffile{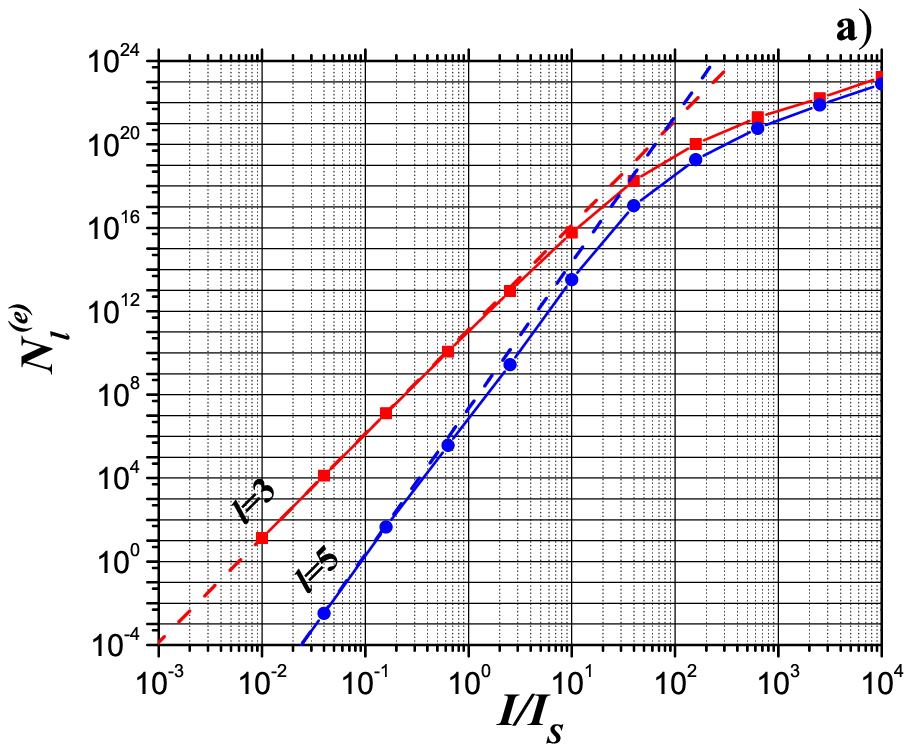}
\epsfxsize6cm\epsffile{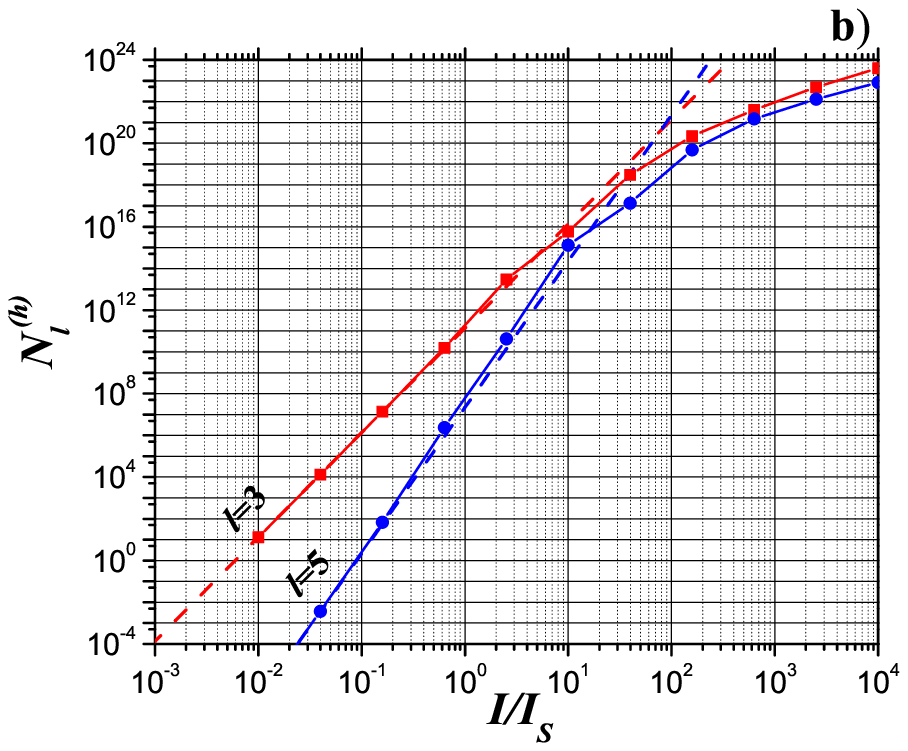} \caption{\small Number of
photons generated per wave period by a a) $e$-polarized beam
($N^e_l$ ) and b) $h$-polarized beam ($N^h_l$ ) versus parameter
$I/I_S$. The solid curves are the results of numerical computation,
dashed lines display the results of perturbation theory,
$\Delta=0.1$.} \label{N1}
\end{figure}

In the region $\vartheta\gtrsim 1$, where the perturbation expansion
(\ref{Lagr_ren}) becomes inapplicable, we have employed the
representation of the Lagrangian density by a convergent series
suggested in Ref.~\cite{CS} and evaluated the integrals in
Eqs.~(\ref{sol_Maxwell3}), (\ref{result}) numerically. The results
of numerical calculations are shown in Fig.~\ref{N1}.
One can see
that, at $\vartheta$ close to unity, the perturbation theory
overstates the number of emitted photons. One can see also the
tendency towards equalization of the number of emitted photons in
neighboring harmonics when $\vartheta$ grows.

It is worth noting that the probability of the third harmonic
generation in the considered case is several orders larger than
that of the second harmonic generation in the crossed beam setup
\cite{HG_st_w}. This can be explained by the effect of stimulated
emission of the photon. To make certain of that, we first notice
that both processes are described by topologically equivalent
Feynman diagrams depicted in Fig.~1b. The only difference consists
in the fact that one of the outgoing photons in our case is a
laser photon, while in the case considered in Ref.~\cite{HG_st_w}
both photons are extraneous. Consider the effect of the third
harmonic generation by two pairs of laser photons colliding at an
angle $2\psi$, $\psi<\Delta$. At the reference frame $K'$, where
the collision is head-on, the probability of our process per unit
time and unit volume can be estimated using Eq.~(4.38) of
Ref.~\cite{HG_st_w}. Since the probability is invariant, we have
in the laboratory frame
\begin{equation}\label{st em}
\frac{dN^{(E-H)}_2}{dVdt}\sim \alpha^2\omega'^4
\bigg(\frac{E_0^{'}}{E_S}\bigg)^8dn_{\psi}\,.
\end{equation}
Here $\omega'$ and $E_0'$ are the frequency and the peak field
strength of the laser at the $K'$-frame. The velocity of the
$K'$-frame is $v\sim\cos\psi\sim 1-\psi^2/2$, and the
corresponding Lorentz factor $\gamma\sim1/\psi$. Hence,
$\omega'=\omega\psi$ and $E_0'=E_0\psi$. The factor $dn_\psi$
arises due to the effect of stimulated photon emission, and has
the meaning of the number of laser photons propagating at the
interval of angles from $\psi$ to $\psi+d\psi$ relative to the
direction of propagation of the laser beam. It is easy to see
using (\ref{Pulse}) that $dn_\psi\sim
|\vec{a}(\vec{k}_\perp)|^2dk_{\perp}^2$ with $k_\perp=\omega\psi$,
and $\vec{a}(\vec{k}_\perp)$ given by Eq.~(\ref{Pulse1}). After
performing integration over $\psi$ in (\ref{st em}), and
multiplying the result by the 4-volume of the focal region
$V_fT\sim1/(\omega\Delta)^4$ we arrive, up to a numerical factor,
to Eq.~(\ref{main_res}) with $l=3$.

In the present letter we have studied the effect of harmonics
generation by an intense focused laser pulse in vacuum. We have
shown that for the laser pulse with $\lambda=1\mu\rm m$ and
$\Delta=0.1$ the number of generated photons reaches the value of
one photon per period at intensity $I\approx5\cdot10^{27}{\rm
W/cm}^2\approx10^{-2}I_S$. The corresponding peak value of the
electric field is one order of magnitude less than the
characteristic QED field $E_S$. This is explained by a very large
value of the effective focal region as compared to the
characteristic Compton 4-volume, $l_C^4/c$. It is very important
that the rates of photon generation for $e$- and $h$-polarized waves
are close to each other and reach an observable level at
$\vartheta\approx 10^{-2}$. It follows from Ref.~\cite{Nar1} that
the effect of pair creation by a focused $e$-polarized laser pulse
arises at the same values of $\vartheta$, while in an $h$-polarized
wave production of pairs begins noticeably later. This means that in
an $h$-polarized wave we will not have the background of secondary
photons radiated by created particles, and hence, just such waves
should be employed for observation of the effect of harmonics
generation by a focused laser pulse in vacuum.

We thank N.L. Manakov for an interesting comment, and the anonymous
referee who suggested the idea for explanation of unexpectedly large
value of the considered effect. This work was supported by the RFBR
grant 06-02-17370-a, by the RF Ministry of Science and Education
(grant RNP 2.11.1972), and by the Russian Federation President
grants MK-2279.2005.2 and NSh-320.2006.2.

\end{document}